\documentclass[%
 reprint,
 amsmath,amssymb,
 aps,
]{revtex4-2}
\usepackage{xcolor}
\usepackage{hyperref}
\hypersetup{
  colorlinks,
  citecolor=blue,
  linkcolor=blue,
  urlcolor=black}

\usepackage{booktabs}
\usepackage{gensymb}
\usepackage{graphicx}% Include figure files
\usepackage{dcolumn}% Align table columns on decimal point
\usepackage{float}
\usepackage{bm}% bold math

\begin{document}

\preprint{APS/123-QED}

\title{Emergent Zeeman-Resilient Superconductivity Beyond the Spin-Paramagnetic Limit in Ultrathin NiBi$_3$}

\author{Gabriel Sant'ana\textsuperscript{1,2}}
\email{gabriel.santana.dasilva@outlook.com}
\author{Leonardo Pessoa da Silva\textsuperscript{1}}
\author{David Möckli\textsuperscript{1}}
\author{Pedro Schio\textsuperscript{3}}
\author{Jan Aarts\textsuperscript{2}}
%\author{David Mockli\textsuperscript{1}}
\author{Kaveh Lahabi\textsuperscript{2}}
\author{Milton A. Tumelero\textsuperscript{1}}
\email{matumelero@if.ufrgs.br}
\space
\affiliation{\textsuperscript{1}Instituto de F\'{i}sica, Universidade Federal do Rio Grande do Sul, 91501-970 Porto Alegre, Brazil.\\
\textsuperscript{2}Kamerlingh Onnes-Huygens Laboratory, Leiden University, P.O. Box 9504, 2300 RA Leiden, The Netherlands.\\
\textsuperscript{3}Centro Nacional de Pesquisa em Energia e Materiais, CP 6192, 13083-970 Campinas, SP, Brazil.}

\date{\today}

\begin{abstract}

The spin-paramagnetic limit sets a fundamental magnetic-field bound for conventional superconductors. Here we show that ultrathin NiBi$_3$ films develop a highly field-resilient superconducting state, with in-plane critical fields surpassing the spin-paramagnetic limit even above 0.9T$_C$. This enhancement is activated by dimensional confinement and depends sensitively on film thickness and morphology. Standard mechanisms, including strong spin-orbit coupling and multiband superconductivity, fail to quantitatively explain the observed robustness. These findings uncover an unconventional pathway for Zeeman-resistant superconductivity in low-dimensional materials beyond known Ising and Rashba scenarios, and further support earlier theoretical predictions of triplet pairing in low-dimensional NiBi$_3$.

\end{abstract}
\maketitle

The robustness of superconductivity against magnetic fields provides key insight into the underlying pairing mechanism. In conventional weak-coupling spin-singlet Bardeen–Cooper–Schrieffer (BCS) superconductors, Zeeman splitting breaks Cooper pairs, leading to the spin-paramagnetic limit (SPL), $B_P = \Delta/\sqrt{2}\mu_B \approx 1.86,T_C$ at zero temperature \cite{clogston1962upper,chandrasekhar1962note}. In most superconductors, however, this limit is masked by orbital depairing, which suppresses superconductivity at lower magnetic fields. In low-dimensional systems, geometric confinement suppresses orbital effects for in-plane magnetic fields, allowing the paramagnetic limit to dominate \cite{tinkham1963effect}. In this regime, several mechanisms can enhance the upper critical field. Strong spin–orbit coupling (SOC) can lock electron spins and reduce Zeeman pair breaking, as observed in transition-metal dichalcogenides (TMDs) \cite{lu2015evidence,saito2016superconductivity,xi2016ising,zeng2018gate,wan2023orbital,xing2017ising,liu2020type} and interfacial superconductors \cite{monteiro2017two,liu2018interface}. Alternatively, unconventional pairing states can also reduce the sensitivity to magnetic fields, particularly spin-triplet superconductivity \cite{ran2019nearly,yang2021spin,huy2008unusual}. In addition, spin-triplet correlations can be generated extrinsically at interfaces through spin-mixing processes, which require the breaking of spin-rotation symmetry, for instance by magnetic interactions or spin–orbit coupling \cite{bergeret2005odd,anwar2010long}.

Bi/Ni bilayers have recently attracted considerable attention as a platform for unconventional superconductivity. In thin epitaxial samples, experimental signatures consistent with chiral $p$-wave-like and spin-triplet pairing have been reported \cite{wang2017anomalous,chauhan2019nodeless,gong2017time,cai2023nonreciprocal}. However, no clear violation of the spin-paramagnetic limit has been established to date.
In contrast, several studies have associated the observed superconductivity with the formation of the intermetallic compound NiBi$_3$ at the interface, with $T_C \sim 4$ K \cite{silva2013superconductivity,vaughan2020origin,park2024superconducting}. While bulk NiBi$_3$ is commonly regarded as a conventional $s$-wave superconductor, it also exhibits signatures of ferromagnetic spin fluctuations and anomalous transport behavior near $T_C$, suggesting a more complex underlying electronic state \cite{herrmannsdorfer2011structure,zhu2012surface,sant2026linear}. Although spin-triplet superconductivity in the bulk has largely been ruled out \cite{zhao2018singlet,shang2023fully}, NiBi$_3$ thin films appear to be strongly influenced by spin–orbit coupling, behaving as a two-dimensional Rashba superconductor with nonreciprocal transport \cite{hayashi2024two,cai2023nonreciprocal}.
Moreover, recent studies indicate that film thickness plays a crucial role in the unconventional behavior of this system \cite{sant2024ni}. In addition, theoretical models suggest that the interplay between exchange fields and strong spin–orbit coupling can generate triplet correlations \cite{chao2019superconductivity}. Nevertheless, it remains unclear whether reduced dimensionality can induce unconventional pairing channels or modify the pair-breaking mechanisms governing the upper critical field.

\begin{figure*}
    \centering
    \includegraphics[width=1\linewidth]{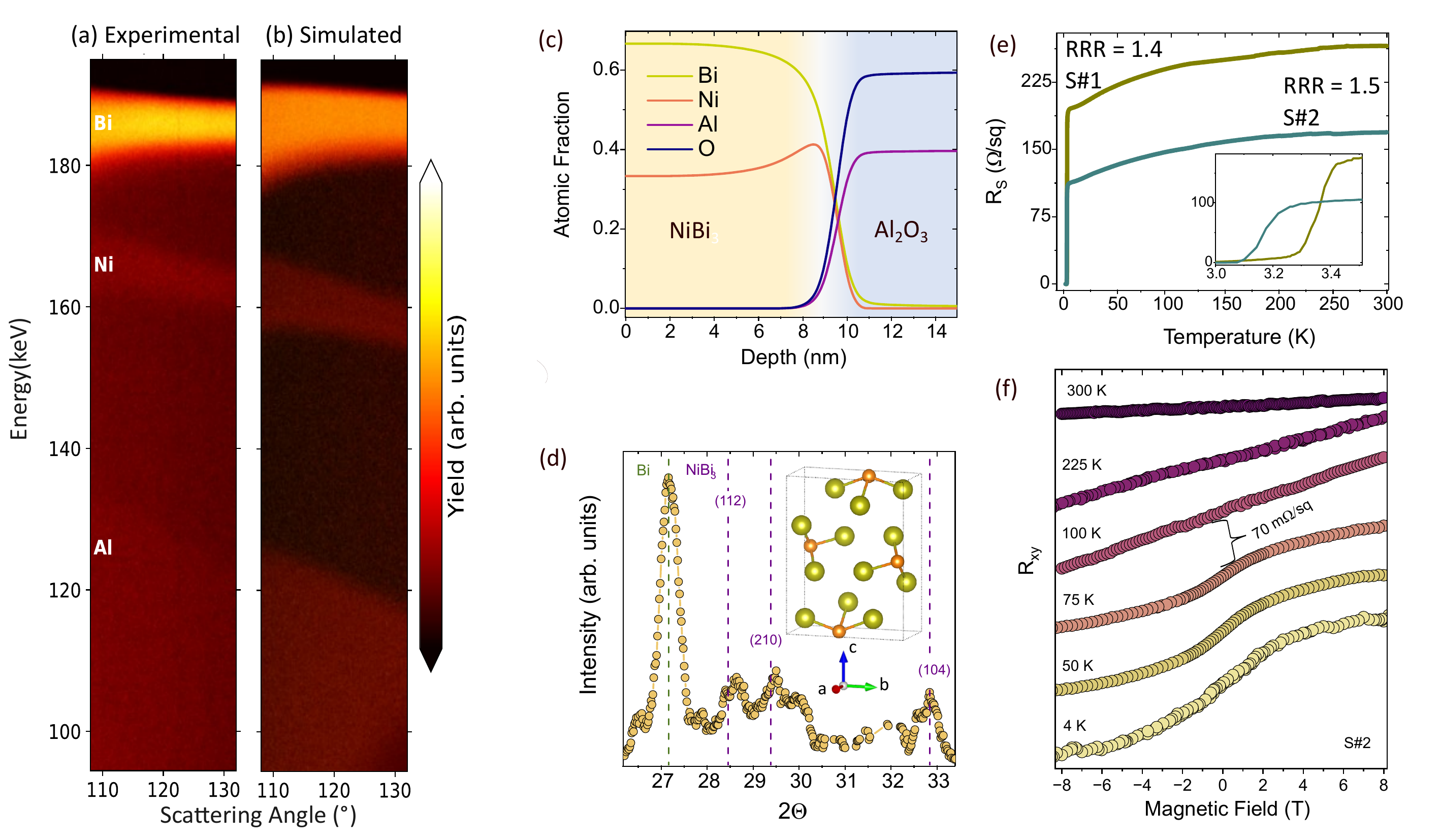}
    \caption{(a) Experimental energy-angle map of 200 keV He$^+$ backscattered ions obtained by MEIS for sample S$\#2$. (b) Monte Carlo simulation using PowerMEIS reproducing the scattering features. (c) Atomic depth profile showing the elemental composition in atomic fraction. (d) Grazing Angle XRD of sample S$\#$2. Purple dashed-line represents  \textit{Inset:} Orthorhombic structure of NiBi$_3$ unit cell. (e) Sheet resistance as function of temperature for S$\#$1 and S$\#$2. Residual resistivity ratio are labeled in the graph. \textit{Inset:} zoom-in region of the superconducting transition. (f) Transversal resistance as function of magnetic field for S$\#$2 at several temperatures.}
    \label{fig1}
\end{figure*}

Here, we report a highly robust superconducting state against magnetic fields in a 6 nm thick NiBi$_3$ film. The thin layer behaves as a two-dimensional superconductor, exhibiting pronounced anisotropy in the upper critical field and exceeding $B_P$ by more than a factor of 1.5 under in-plane magnetic fields, thereby violating the spin-paramagnetic limit at temperatures close to the critical point, up to 0.9T$_C$. Standard models based on strong coupling, spin–orbit scattering, multiband effects, Rashba coupling, and Ising-type spin–orbit coupling fail to account for the observed behavior. Our results instead indicate an emergent pair-breaking regime driven by the interplay of reduced dimensionality, spin–orbit coupling, and magnetic fluctuations. These findings establish low-dimensional NiBi$_3$ as a new platform for superconductivity beyond established Ising and Rashba paradigms, pointing to an unconventional mechanism. 

The films were grown by molecular beam epitaxy (MBE) at 300 K throught the deposition of a Bi (10nm)/Ni (1nm) bilayers, on top of Al$_2$O$_3$ (0001) and MgO (001) single-crystal substrates. The base pressure of $2 \times 10^{-10}$ mbar. Reflection high-energy electron diffraction (RHEED) patterns confirmed a layer-by-layer growth mode for the initial five monolayers of Ni. Structural, morphological and compositional analyses were complemented by using Grazing Incident angle X-Ray Diffraction (GIXRD) and Medium-Energy Ion Scattering (MEIS), which provides sub-nanometer depth resolution for thin-film characterization. Full characterization data are provided in the Supplementary Material (SM). %MEIS measurements were performed using a 200 keV He$^+$ beam, and the resulting energy-angle map of the backscattered ions is shown in Fig. \ref{fig1}{\color{blue}a}. Monte Carlo simulations consistent with the experimental data, performed using the PowerMEIS software \cite{marmitt_powermeis_2025}, are shown in Fig. \ref{fig1}{\color{blue}b}. The optimized depth profile, displayed in Fig. \ref{fig1}{\color{blue}c}, reveals a 6.5 nm NiBi$_3$ layer beneath a 1 nm BiNi interface layer. The formation of NiBi$_3$ was further confirmed by GIXRD, which shows Bragg peaks consistent with the orthorhombic \textit{Pnma} structure. 

Fig. \ref{fig1}{\color{blue}a} shows the experimental energy–angle map obtained from MEIS measurements. Monte Carlo simulations using the PowerMEIS software reproduce the main scattering features (Fig. \ref{fig1}{\color{blue}b}). The complete Rutherford Backscattering Spectrum at distinct angles are provided in the SM. The clear step in the analysis indicate the flat layer. The resulting compositional depth profile, shown in Fig. \ref{fig1}{\color{blue}c}, reveals the formation of a $\sim$ 6 nm NiBi$_3$ instead of a Bi/Ni bilayer, as nominally deposited. In between the NiBi$_3$ layer and the Al$_2$O$_3$ substrate, there is small interface layer, with about 1 nm, rich in Ni. The whole resolution of this MEIS profiling analysis is about 1 nm, which again confirms the formation of a flat and stoichiometric NiBi$_3$ layer. GIXRD measurement, displayed in the Figure \ref{fig1}{\color{blue}d} further indicates the presence of the \textit{Pnma} orthorhombic NiBi$_3$ phase, through the indexed diffraction pattern highlighted by the purple-dashed line. Schematic unit cell of NiBi$_3$ are inset in the Fig. \ref{fig1}{\color{blue}d}. The structural characterization for the samples grown over the MgO substrate is presented in SM.

Fig. \ref{fig1}{\color{blue}e} presents the zero-field electrical transport measurements of two samples grown on Al$_2$O$_3$. The residual resistivity ratio (RRR), determined as RRR = R(300 K)/R(5 K), was estimated to be approximately 1.50, in agreement with prior reports on Bi/Ni thin films \cite{park2024superconducting}. The similar RRR values between samples suggest comparable impurity content and microstructural properties. The insets in Fig. \ref{fig1}{\color{blue}e} display the superconducting transitions, with T$_C$ of 3.36 K and 3.25 K for S$\#1$ and S$\#2$, respectively, defined as the 50$\%$ drop from the normal-state resistance. Both samples exhibit sharp transitions ($\Delta T_C$ $<$ 0.20 K), indicating a homogeneous superconducting state. 

Furthermore, transverse resistance measurements of sample S$\#2$ were performed over a wide temperature range, as shown in Fig. \ref{fig1}{\color{blue}e}. From 75 K down to the vicinity of the superconducting transition, a nonlinear low-field contribution resembling an anomalous Hall effect is observed. Above $\sim$100 K, this contribution vanishes and the response is dominated by the ordinary Hall effect. Similar behavior has been reported in non-epitaxial Bi/Ni bilayers \cite{hayashi2024two} and, more recently, in high-quality NiBi$_3$ single crystals, where it was attributed to ferromagnetic spin fluctuations arising from the combined effects of skew scattering and the absence of long-range magnetic order \cite{sant2026linear}. Notably, this anomalous contribution becomes prominent close to the superconducting transition. Consistent with the MEIS analysis, which shows no residual Ni layer in our samples, the observed Hall response cannot originate from a ferromagnetic Ni contribution. Instead, it is most naturally attributed to an intrinsic property of the NiBi$_3$ layer, in agreement with the spin-fluctuation scenario reported for bulk single crystals.

\begin{figure*}
    \centering
    \includegraphics[width=1\linewidth]{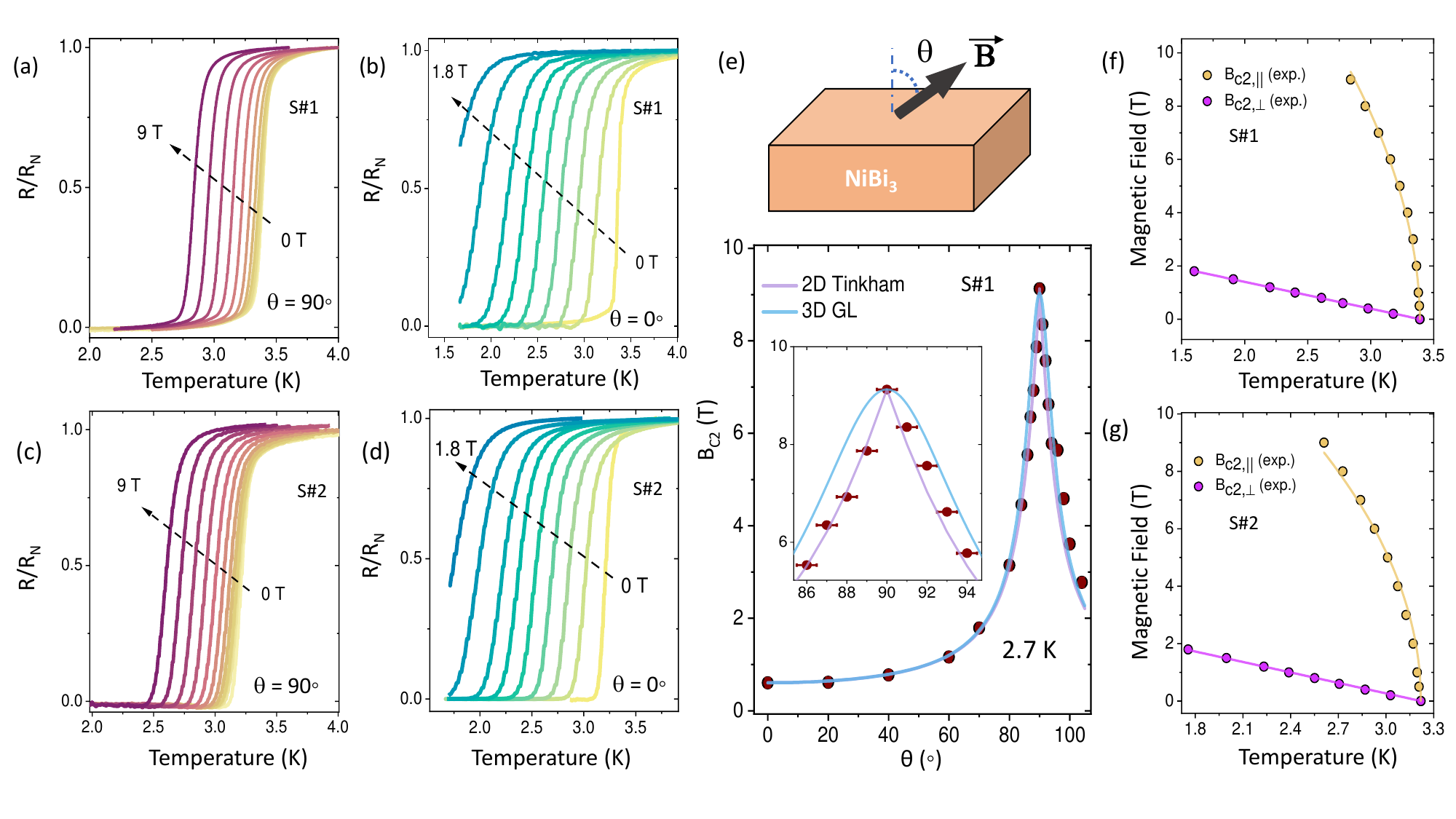}
    \caption{Magnetotransport measurements in parallel and perpendicular field configurations for Bi(10 nm)/Ni(1 nm)/Al$_2$O$_3$: (a), (b) Sample S$\#$1; (c),   (d) Sample S$\#$2. (e) Angular dependence of the upper critical field, B$_{C2}$ ($\theta$), where $\theta$ is the angle between the magnetic field and the      direction perpendicular to the Bi/Ni surface, as shown in the upper panel. (f), (g) Upper critical field as a function of temperature. The red and black solid lines correspond to the anisotropy theory of Ginzburg-Landau (Eqs. \ref{eqn:gl_perp} and \ref{eqn:gl_parall}, respectively).}
    \label{fig:RxTxB}
\end{figure*}

To further investigate the superconducting properties, we performed magnetotransport measurements under in-plane ($\theta$ = 90$\degree$) and out-of-plane ($\theta$ = 0$\degree$) magnetic fields. As shown in Fig. \ref{fig:RxTxB}{\color{blue}a-d}, the superconducting transition systematically shifts to lower temperatures with increasing field strength. A pronounced anisotropy is observed: superconductivity is fully suppressed by an out-of-plane field of 2 T, whereas it remains remarkably robust in the in-plane configuration, with only an 80\% reduction of T$_C$ at 9 T. To elucidate this anisotropic behavior, we extract the B$_{C2}$ at 2.7 K as a function of the field angle, shown in Fig. \ref{fig:RxTxB}{\color{blue}e}. The data follow the characteristic cusp-like dependence described by the 2D Tinkham model \cite{tinkham1963effect}, given by $(B_{C2}(\theta)sin\theta/B_{C2}^{||})^{2} + |B_{C2}(\theta)cos\theta/B_{C2}^{\perp}| = 1$, confirming the two-dimensional nature of superconductivity in the NiBi$_3$ layer.

The temperature-dependent upper critical fields in out-of-plane (B$_{C2}^{\perp}$) and in-plane (B$_{C2}^{||}$) directions are plotted in Fig. \ref{fig:RxTxB}{\color{blue}f,g}. In the former case, the transition from the normal state to a superconducting vortex phase is describe by the anisotorpy Ginzburg Landau (AGL) expression 

\begin{equation}
    B_{c2}^{\perp}(T) = \frac{\phi_{o}}{2\pi\xi_{||}^{2}}\left(1 - \frac{T}{T_C}\right) \label{eqn:gl_perp},
\end{equation}
being $\phi_{o}$ the magnetic flux quantum and $\xi_{||}$ the in-plane coherence length, respectively. From the magenta line fits, we extract $\xi_{||}$ as (9.8 $\pm$ 0.1) nm for S\#1 and (9.2 $\pm$ 0.2) nm for S\#2, whose values are similar to other non-epitaxial Bi/Ni bilayers reports \cite{vaughan2020origin,sant2024ni}. 

The response to an in-plane magnetic field further highlights the two-dimensional nature of the superconductivity, exhibiting an early curvature in the phase diagram. The goldenrod color lines correspond to the 2D AGL equation:
\begin{equation}
    B_{c2,}^{||}(T) = \frac{\sqrt{12}\phi_{o}}{2\pi\xi_{||}d_{sc}}\sqrt{1 - \frac{T}{T_C}} \label{eqn:gl_parall},
\end{equation}
where d$_{sc}$ represents the effective superconducting thickness. The best fit yields d$_{sc}$ = 5.1 nm for S\#1  and  6.2 nm for S\#2, whose values matches very well MEIS analysis for S$\#2$ ($\sim$ 6 nm). This consistency strongly suggest that the critical field behavior can be well captured by an orbital-limited description, with superconductivity confined to the NiBi$_3$ layer. The upper critical field data for the samples deposited over MgO substrate is presented in the SM.

\begin{figure}[h]
    \centering
    \includegraphics[width=1\linewidth]{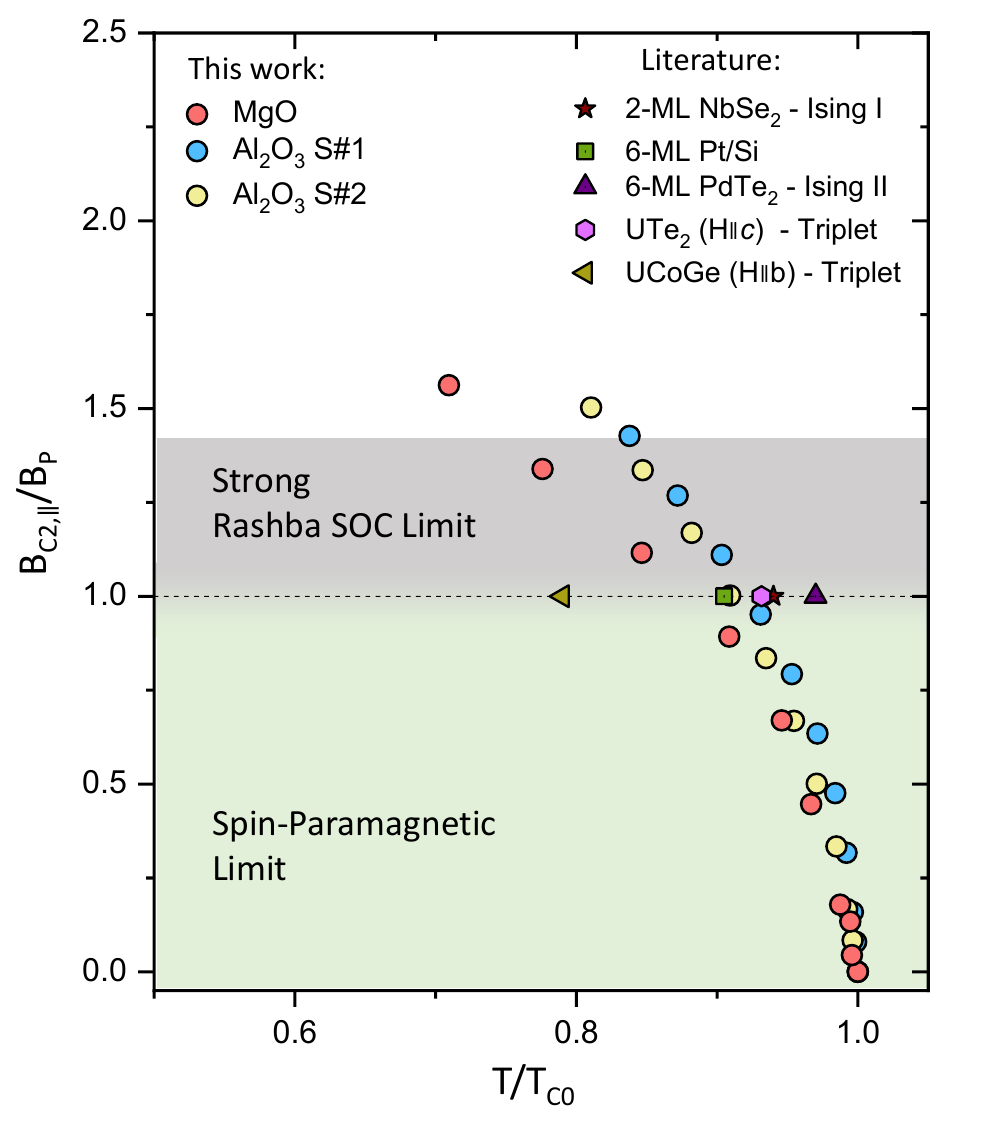}
    \caption{In-plane upper critical field divided by B$_P$ as function of reduced temperature for S\#1, S\#2 and MgO. Literature data for NbSe$_2$ \cite{xi2016ising}, PdTe$_2$ \cite{liu2020type}, UTe$_2$ \cite{ran2019nearly}, UCoGe \cite{huy2008unusual} and epitaxial Pt growth on striped incommensurate (SIC) Si(111) \cite{liu2018interface}, were extracted from previous reports for comparison. }
    \label{fig:Bc2_bp_t}
\end{figure}

In spite of the orbital scenario accounting for the temperature-dependent scaling of $B_{C2}^{\parallel}$, the data reveal a striking feature: the conventional spin-paramagnetic limit is systematically violated. As highlighted by the green-shaded region in Fig. \ref{fig:Bc2_bp_t}, all samples exceed such a limit with $B_{C2}^{\parallel}/B_P \approx 1.5$, corresponding to $B_P$ values of 6.2 T, 6.1 T, and 4.5 T for samples S$\#$1, S$\#$2, and MgO, respectively. 

Some known mechanisms can, in principle, enhance the spin-paramagnetic limit in conventional superconductors. In the strong electron–phonon coupling regime, Eliashberg theory predicts a renormalization of the Pauli limit, which is commonly written, within standard approximations, as $B_P^{s-c.} \propto (1+\lambda_{ep})^{1/2} B_P$, where $\lambda_{ep}$ is the electron–phonon coupling constant \cite{orlando1979critical}. NiBi$_3$ has been discussed as a weak-coupling superconductor (2$\Delta$/k$_B$T$_C$ $\sim$ 3.51) \cite{zhao2018singlet} and the $\lambda_{ep}$ constant has been reported as $\sim 1.1$ \cite{fujimori2000superconducting}, yielding an upper bound of $B_P^* \approx 1.45\,B_P$ at zero temperature, whose value lies below our experimental data. Therefore, corrections from strong-coupling superconducting gap energy renormalization cannot account for the enhanced robustness of superconductivity against magnetic fields in quasi-2D NiBi$_3$. %\textcolor{red}{Additionally, we argue that any bulk-like effect could account for violation of the of the SPL in these system, such as multi-orbital superconductivity, since it would be observed in other thin films, such as to ones prepared with magnetron sputtering and discussed in the SM \cite{suppmat}.}

Moreover, in disordered superconductors with SOC, nonmagnetic impurities and structural defects can induce spin-dependent scattering, which randomizes the electron spins orientations and thereby weakens Zeeman pair breaking. In two-dimensional systems, this effect is captured by the Klemm–Luther–Beasley (KLB) theory \cite{klemm1975theory}, where the spin-orbit scattering (SOS) should be the dominant process, i.e. $\tau_{tot}\approx\tau_{SO}$ to substantially increase the SPL. Nevertheless, reproducing the observed violation of the spin-paramagnetic limit within the KLB framework would require a quite short spin–orbit scattering time of $\tau_{SO}\sim10^{-14}$ s (see SM). This value is substantially smaller than the independently estimated $\tau_{tot}\sim10^{-12}$–$10^{-13}$ s for these samples. This indicates that SOS and impurity disorder cannot account for the observed SPL violation.

Alternatively, Rashba SOC is known to induce mixing between singlet and triplet states, which naturally weakens the paramagnetic effect \cite{gor2001superconducting}. In heterostructure, where structural asymmetric potential is present perpendicular to the substrate plane, Rashba SOC is expected to produces in-plane spin-orbit magnetic field. In particular, for 2D conventional superconductor with strong Rashba SOC, the paramagnetic limit at zero temperature is modified as B$_P^* \sim \sqrt{2}B_P$ \cite{liu2018interface,yoshizawa2021atomic}. 
%Indeed, previous reports have argues in favor of Rashba-type SOC in Bi/Ni bilayers through nonreciprocal transport measurements \cite{hayashi2024two,cai2023nonreciprocal}. 
Nonetheless, the upper bound from B$_P^*$ alone fails to describe for the experimental data, as represented in the gray-shaded region in Fig. \ref{fig:Bc2_bp_t}.

Large in-plane upper critical fields ($B_{C2}^{\parallel}$) have been widely reported in two-dimensional superconductors exhibiting Zeeman-protected superconductivity, where SOC locks electron spins out of plane and suppresses paramagnetic pair breaking \cite{lu2015evidence,saito2016superconductivity,zeng2018gate}. In TMDs, this effect originates from the intrinsic lack of inversion symmetry in the crystal structure, leading to valley-dependent spin splitting at the K (K$'$) points \cite{xing2017ising,xi2016ising}. This mechanism is commonly referred as type-I Ising superconductivity. Similar Zeeman-protected behavior has also been observed in ultrathin centrosymmetric Pb films, where interface-induced in-plane inversion symmetry breaking at reconstructed Pb/Si(111) surfaces \cite{liu2018interface}. However, such Ising-type mechanisms are unlikely to account for our observations. First, NiBi$_3$ crystallizes in the centrosymmetric orthorhombic \textit{Pnma} structure (Fig. \ref{fig1}{\color{blue}d}). Moreover, the interface-induced in-plane inversion symmetry breaking appears unlikely to play a dominant role in our samples once the NiBi$_3$ thickness (5–6 nm) exceeds the typical decay length of interface-induced SOC, which is generally limited to only a few monolayers \cite{yoshizawa2021atomic}. 

More recently, type-II Ising superconductivity has been proposed for centrosymmetric systems, where strong spin-orbit coupling acting on symmetry-protected multi-orbital degeneracies leads to spin-orbital locking and an effective out-of-plane spin polarization that protects Cooper pairs against in-plane magnetic fields \cite{liu2020type,wang2019type}. Although NiBi$_3$ hosts strong SOC and a multi-orbital electronic structure, this mechanism relies on well-defined crystalline symmetry to preserve orbital degeneracies and a coherent spin-orbital locking across momentum space. In these polycrystalline films the local spin–orbit fields are randomly oriented across different grains, whose vector contributions average out, suppressing any net out-of-plane spin–momentum locking.

Finally, a plausible mechanism for the anomalously large $B_{C2}^{\parallel}$ and the persistence of superconductivity beyond the spin-paramagnetic limit up to 0.9$T_C$ is the emergence of spin-triplet correlations. In equal-spin triplet states, paramagnetic pair breaking can be strongly suppressed, such that the upper critical field becomes primarily limited by orbital effects \cite{ran2019nearly,slooten2009enhancement}. As discussed above, our analysis within the anisotropic Ginzburg–Landau framework indicates that pair breaking in this regime is predominantly orbital. Spin-triplet superconductivity may arise in two symmetry classes: odd-parity and odd-frequency \cite{berezinskii1974new,bergeret2005odd}. While odd-parity triplet states are generally sensitive to impurity scattering and therefore unlikely in disordered polycrystalline films, odd-frequency triplet correlations can emerge from singlet pairs through spin mixing in the presence of broken spin-rotational symmetry \cite{eschrig2008triplet}.

Apart from the SPL, NiBi$_{3}$ presents additional key ingredients for the triplet scenario, such as ferromagnetic spin fluctuations, that have been predicted to mediate pairing with triplet excitations \cite{hattori2012superconductivity,kitamura2024spin,kreisel2022spin,yang2021spin}. Indeed, Fig. \ref{fig1}{\color{blue}f} displays transverse resistance measurements on sample S$\#$2 at 4 K, revealing a nonlinear field dependence just above $T_C$. Such nonlinear Hall behavior has been discussed in terms of spin fluctuations \cite{sant2026linear}. These spin fluctuations near $T_C$ may act as a dynamic exchange field that breaks spin-rotational symmetry and can potentially promote triplet correlations \cite{zyuzin2022fluctuation}. The presence of a strong Rashba SOC \cite{cai2023nonreciprocal,hayashi2024two} could polarize the triplet spin vector into the plane of the layer \cite{PhysRevB.2024.RashbaSuperconductor,frigeri2004superconductivity}.Furthermore, this interpretation aligns with previous studies on the Bi/Ni system that observed signatures of triplet pairing \cite{gong2015possible,chao2019superconductivity,gong2017time, wang2017anomalous}, particularly in theoretical models accounting for the interplay between exchange fields, SOC and superconductivity.

The dimensionality of the NiBi$_3$ layer appears to play a decisive role in the emergence of unconventional superconductivity. Systematic measurements were performed on Bi(10$\cdot$ x nm)/Ni(x nm) bilayers with $x$ = 1, 2, and 3, grown by magnetron sputtering (see SM). For the thinnest specimen ($x$ = 1), no signature of NiBi$_3$ formation was detected in either structural or transport measurements, likely due to the island-like growth of ultrathin Ni, which prevents the formation of a continuous superconductor layer. In contrast, for thicker Ni layers ($x$ = 2 and 3), the formation of a continuous NiBi$_3$ layer, displaying a 3D superconducting response. In this regime, the $B_{C2}^{\parallel}$ lies below the spin paramagnetic limit and is well captured by a conventional superconducting model with minor spin–orbit contributions. Such a result is consistent with previous finding, where NiBi$_3$ size confinement structures shows large upper critical field beyond the 3D counterpart \cite{herrmannsdorfer2011structure}.

Overall, NiBi$_3$ emerges as a rich platform for correlated quantum phenomena, hosting superconductivity, complex electronic structure with strong SOC, non-trivial band topology \cite{adriano2023bulk,zhang2025nodal} and exotic magnetic phenomena \cite{sant2026linear,zhu2012surface,herrmannsdorfer2011structure,Rollano_2023}. In this context, the observation of a systematic violation of the SPL, reaching $B_{C2}^{\parallel}/B_P \approx 1.5$ at reduced temperatures of about $0.9T_C$, is consistent with a superconducting state with enhanced robustness against Zeeman pair breaking. Notably, the temperature evolution of $B_{C2}^{\parallel}/B_P$ closely follows the behavior reported in systems exhibiting unconventional or spin-protected superconductivity, where the paramagnetic limit is violated at $T/T_C \sim 0.8$-$0.9$, as pointed out in Fig. \ref{fig:Bc2_bp_t}.
The robustness against the magnetic field is striking even comparing to 3D Ising superconductor, in which the SPL occurs at about 0.65T$_C$ \cite{Ising3D.2025}. In others multi-band and strong-coupling superconductors the SPL occurs at much lower reduced temperatures \cite{Twobands-Xing2017}.

In summary, we have demonstrated that a 6 nm NiBi$_3$ layer hosts a robust quasi-2D superconducting phase that remains stable against high in-plane magnetic fields, exceeding the spin-paramagnetic limit by a factor of 1.5 at 0.9T$_{C}$. The strong thickness dependence reveals that dimensional confinement plays a decisive role, as the paramagnetic-limit violation emerges only in the ultrathin regime. The enhancement of $B_{C2}^{\parallel}$ was analyzed within standard mechanisms for conventional superconductors, incorporating spin–orbit scattering and spin–momentum locking. However, neither description fully accounts for the experimental data. Instead, our results suggest that the coexistence of strong Rashba-type SOC and ferromagnetic spin fluctuations near T$_C$ can provide a promising route for the conversion of spin-singlet into spin-triplet correlations that are resilient to Zeeman pair breaking. These findings establish low-dimensional NiBi$_3$ as a new platform for exploring fluctuation-assisted triplet superconductivity and unconventional pairing in spin–orbit-coupled quantum materials.

\begin{acknowledgments}
The authors would like to acknowledge the LCN NanoLab
at Physics Institute of UFRGS for the support facilities in this work. The authors also would like to acknowledge the funding agencies CAPES, CNPq and FAPERGS. M.A.T. would like to acknowledge the funding agencies CNPq (Grant No. 313809/2023-2) and FAPERGS (24/2551-12877). M.A.T. also acknowledge the support of the INCT project Advanced Quantum Materials, involving the Brazilian agencies CNPq (Proc. 408766/2024-7). This research used facilities of the Brazilian Nanotechnology National Laboratory (LNNano), part of the Brazilian Centre for Research in Energy and Materials (CNPEM), a private non-profit organization under the supervision of the Brazilian Ministry for Science, Technology, and Innovations (MCTI). The LCIS staff is acknowledged for the assistance during the experiments 20242308 and 20261176.

\end{acknowledgments}

\section*{SUPPLEMENTAL MATERIAL:
Emergent Zeeman-Resilient Superconductivity Beyond the Spin-Paramagnetic Limit in Ultrathin NiBi$_3$}

\section{Samples fabrication and Structural characterization}

\begin{figure*}[h!]
    \centering
    \includegraphics[width=0.8\linewidth]{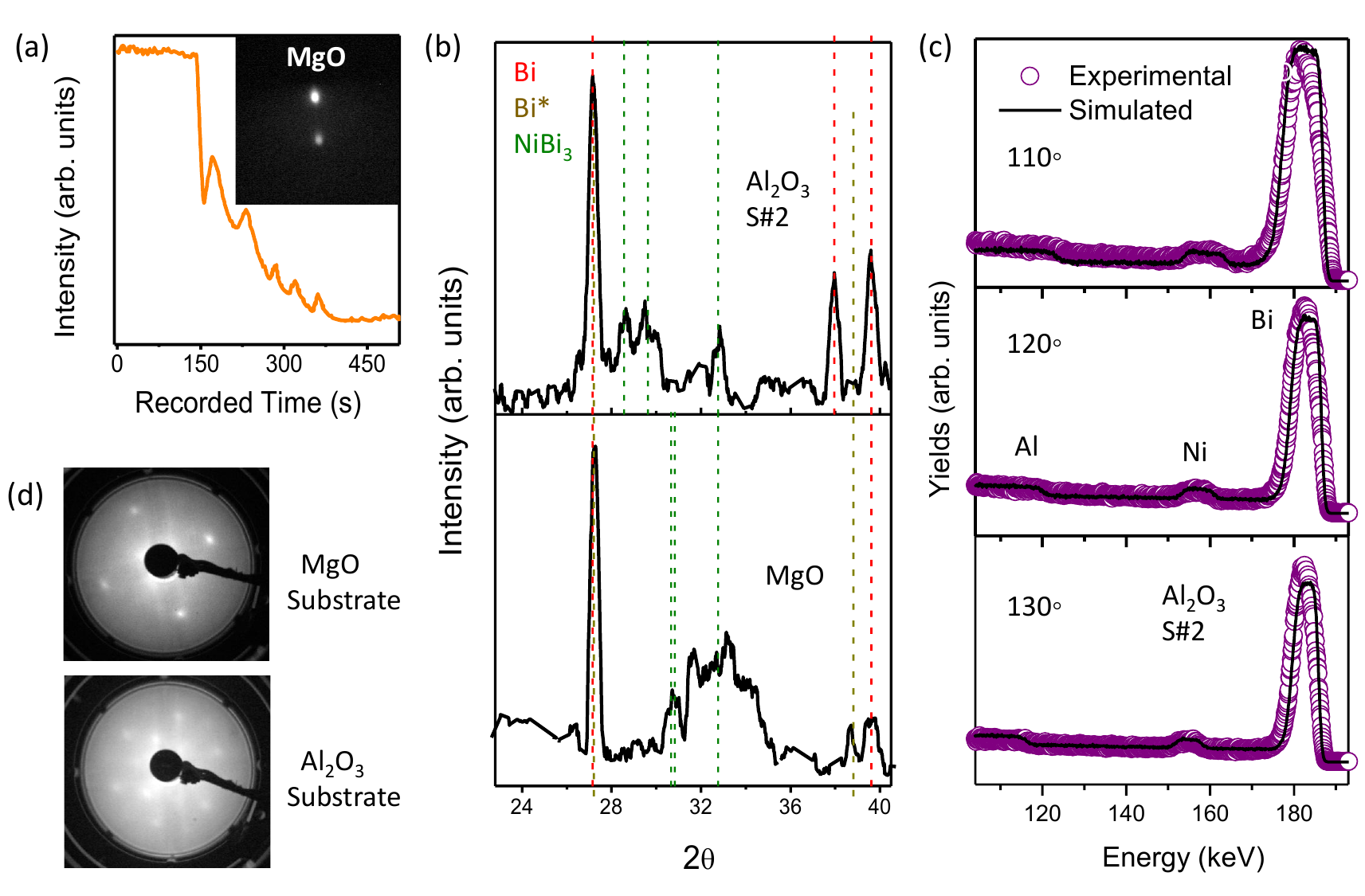}
    \caption{(a) RHEED intensity oscillation for Ni growth on the top of MgO. (b) GIXRD diffractogram of the MBE-samples. Red and dark-yellow dashed lines stems for rhombohedral and cubic Bi structure, while dashed green line refers to NiBi$_{3}$ orthorhombic structure. (c) MEIS spectrum for Bi(10 nm)/Ni(1 nm)/Al$_{2}$O$_{3}$ - S\#2 at room temperature in different incident angle. (d) LEED pattern of the substrates before deposition. }
    \label{fig:characterization}
\end{figure*}

Bi(10 nm)/Ni(1 nm) bilayers were deposited at 300 K via molecular beam epitaxy (MBE) into MgO(001) and Al$_2$O$_3$(001) substrates under a base pressure of $2 \times 10^{-10}$ mbar. Prior to deposition, the substrates were cleaned using a standard RCA procedure. Ni was deposited at a rate of 1.08 monolayers (ML)/min to a total thickness of 5 ML under a chamber pressure of $1.5 \times 10^{-9}$ mbar, followed by Bi deposition at 1 ML/min for 50 ML under $1.8 \times 10^{-8}$ mbar. Reflection High-Energy Electron Diffraction (RHEED) intensity oscillations observed during Ni growth on MgO (Fig. \ref{fig:characterization}\textcolor{blue}{a}) indicate a layer-by-layer growth mode. In contrast, no RHEED oscillations were observed during Bi deposition. Complementary, Low-Energy Electron Diffraction (LEED) pattern of the substrates are presented in Fig. \ref{fig:characterization}{\color{blue}b}. %Atomic Force Microscopy (AFM) images of the resulting films on MgO and Al$_2$O$_3$ (Fig. \ref{fig:characterization}\textcolor{blue}{b,c}) reveal polycrystalline Bi grains. On MgO, the root-mean-square (RMS) roughness was approximately 28 nm, with an average grain width of 110 nm. On Al$_{2}$O$_{3}$, a smoother surface was observed (RMS $\approx$ 25 nm), but with significantly larger grain sizes up to 600 nm.

Fig. \ref{fig:characterization}{\color{blue}c} presents Grazing Incidence X-ray Diffraction (GIXRD) data, obtained using Bruker D8 Advanced with Cu K$\alpha_{1}$ radiation at a fixed incident angle of 1$^{\circ}$. The diffraction patterns reveal Bragg peaks consistent with rhombohedral Bi (space group R3m) indicated by red dashed lines. For films grown on MgO, additional reflections consistent with cubic Bi were also observed (dark-yellow dashed lines). Notably, several mid-intensity peaks corresponding to the orthorhombic NiBi$_{3}$ phase (space group \textit{Pnma}) were identified (green dashed lines), suggesting the formation of intermetallic compounds at the interface. The relatively high background noise in the diffractogram is attributed to the significant surface roughness relative to the film thickness.

Fig.~\ref{fig:characterization}{\color{blue}d} shows the medium-energy ion scattering (MEIS) spectra of sample S$\#$2 measured using a 200 keV He$^+$ beam at different scattering angles (110$^\circ$, 120$^\circ$, and 130$^\circ$). The experimental backscattering yields (open symbols) are plotted as a function of the scattered ion energy and compared with simulated spectra (black solid lines). Distinct peaks associated with backscattering from Bi, Ni, and Al atoms can be clearly identified.

%In the main text, we introduced the total relaxation time ($\tau_{tot.}$) and compared it with the spin-orbit scattering lifetime ($\tau_{SO}$). Here, we detail the determination of the charge carrier density ($n$) and $\tau_{tot.}$. 

%The Hall coefficient ($R_{H}$) was extracted from the linear slope of the Hall voltage and scaled by the NiBi$_{3}$ thickness, yielding $R_{H} = m \times d_{sc} \approx +1.95 \times 10^{-9}\Omega\cdot m/T$. This corresponds to a carrier density of $n = 3.2 \times 10^{27} m^{-3}$, consistent with bulk NiBi$_3$ previous report. 

\section{Magnetotransport in the MgO substrate}

\begin{figure*}[h!]
    \centering
    \includegraphics[width=0.63\linewidth]{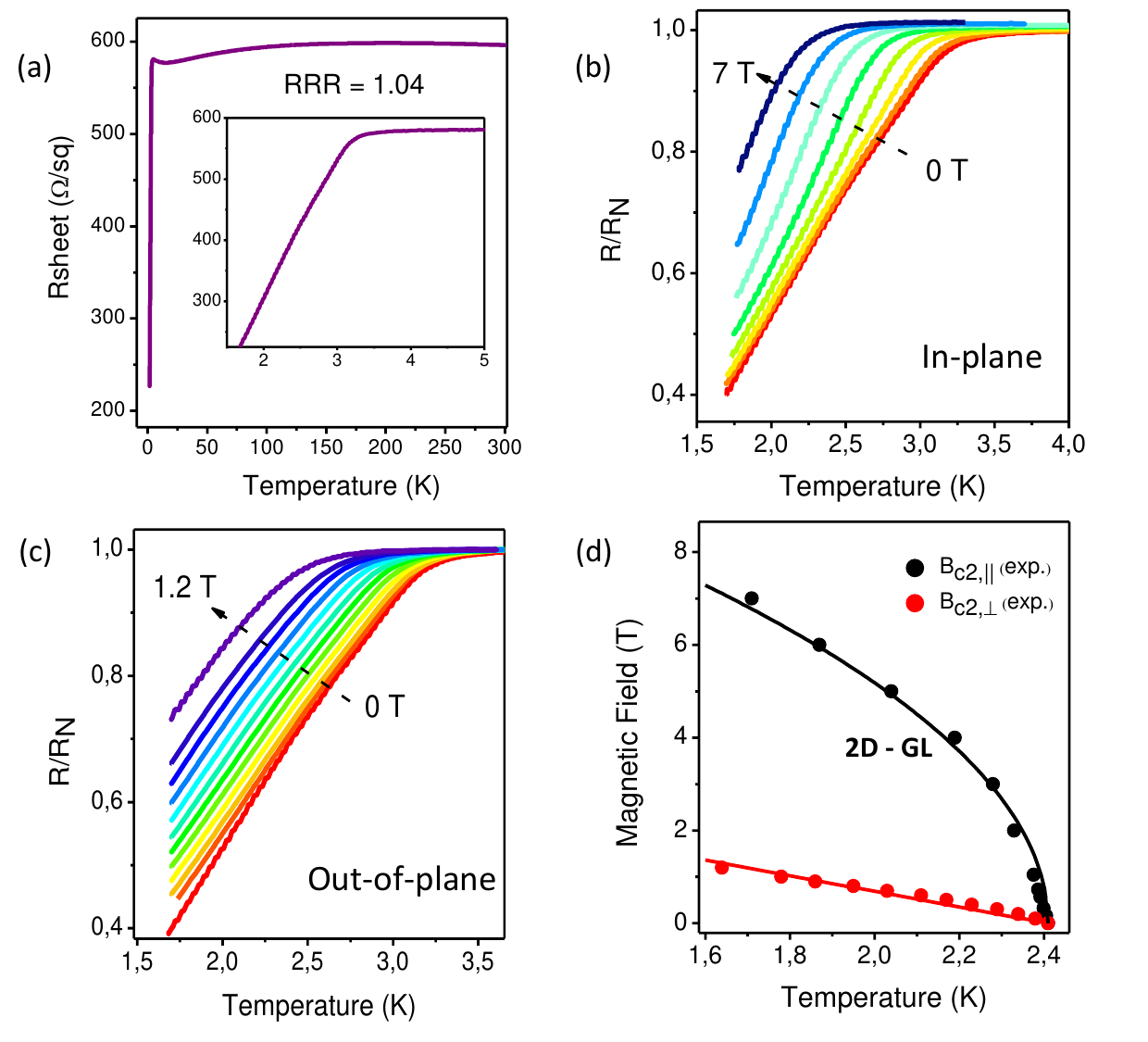}
    \caption{(a) Zero-field resistance vs temperature curve. \textit{Insert:} zoom-in of superconducting transition. Normalized resistance in function of temperature for magnetic field applied (b) in-plane and (c) out-of-plane. B$_{c2}$ vs Temperature phase diagram. Solid lines stems from Anisotropic Ginzburg-Landau (AGL). }
    \label{fig:mgo_data}
\end{figure*}

In the main text, we presented the normalized in-plane upper critical field of Bi (10 nm)/Ni(1 nm)/MgO(001) sample. The complete magnetotransport dataset will be reported here, measured using a PPMS Dynacool 9 T system. The full temperature-dependent resistance curve, shown in Fig. \ref{fig:mgo_data}{\color{blue}a}, displays behavior consistent with the NiBi$_{3}$ intermetallic compound. However, the low residual resistance ratio (RRR $\approx$ 1.04), as compared to the Al$_2$O$_3$-based sample, indicates a significantly more disordered bilayer. This larger disorder can be related to the coexistence of rhombohedral and cubic Bi phases. Since only the rhombohedral Bi phase is known to react with Ni to form the superconducting NiBi$_{3}$ compound, the presence of the cubic phase inhibits uniform intermetallic layer formation. As a result, NiBi$_{3}$ is expected to nucleate in spatially separated domains, instead of forming a continuous superconducting film.

This structural inhomogeneity manifests itself in the transport data. Specifically, a noticeable upturn in resistance appears around 10 K and extends to the onset of superconductivity. Such upturns have been widely associated with inhomogeneous superconductors, where insulating or poorly conducting phases coexist with superconducting ones, resulting in disrupted current paths and increased resistance just above the superconducting transition. Furthermore, the superconducting transition is broad ($\Delta T_c >$ 2 K), which is also consistent with a granular superconducting state. In this scenario, percolation of the superconducting current is suppressed due to spatially disconnected NiBi$_{3}$ grains, likely embedded in a non-superconducting Bi matrix. This behavior is typical of disordered or amorphous superconducting systems and has been discussed in the context of both granular growth and phase-separated films. Consequently, zero resistance is not reached within the accessible temperature range, and we define $T_c$ as the temperature corresponding to 70\% of the normal-state resistance, yielding $T_c = 2.41$ K.

Fig. \ref{fig:mgo_data}{\color{blue}b,c} display the in-plane and out-of-plane magnetotransport, respectively. A strong anisotropy is observed: in the perpendicular field configuration, superconductivity is suppressed for $B_\perp > 1.2$T, whereas the transition remains largely unaffected under parallel magnetic field. The out-of-plane upper critical field $B_{c2}^\perp(T)$ was analyzed within the Ginzburg–Landau framework (Eq. {\color{blue}1} main text), yielding an in-plane coherence length of $\xi_{||} = 9$ nm. The $B_{c2}^{||}$ curve in Fig. \ref{fig:mgo_data}{\color{blue}} reveals clear two-dimensional behavior through the upward curvature near $T_c$. By fitting the parallel-field data using the 2D-AGL (Eq. {\color{blue}2} main text), we estimate the effective superconducting thickness as $d_{sc} \approx 8$ nm.

\section{3D NiBi$_3$ Layer: Sputtering Samples}

\begin{figure*}[h]
    \centering
    \includegraphics[width=0.78\linewidth]{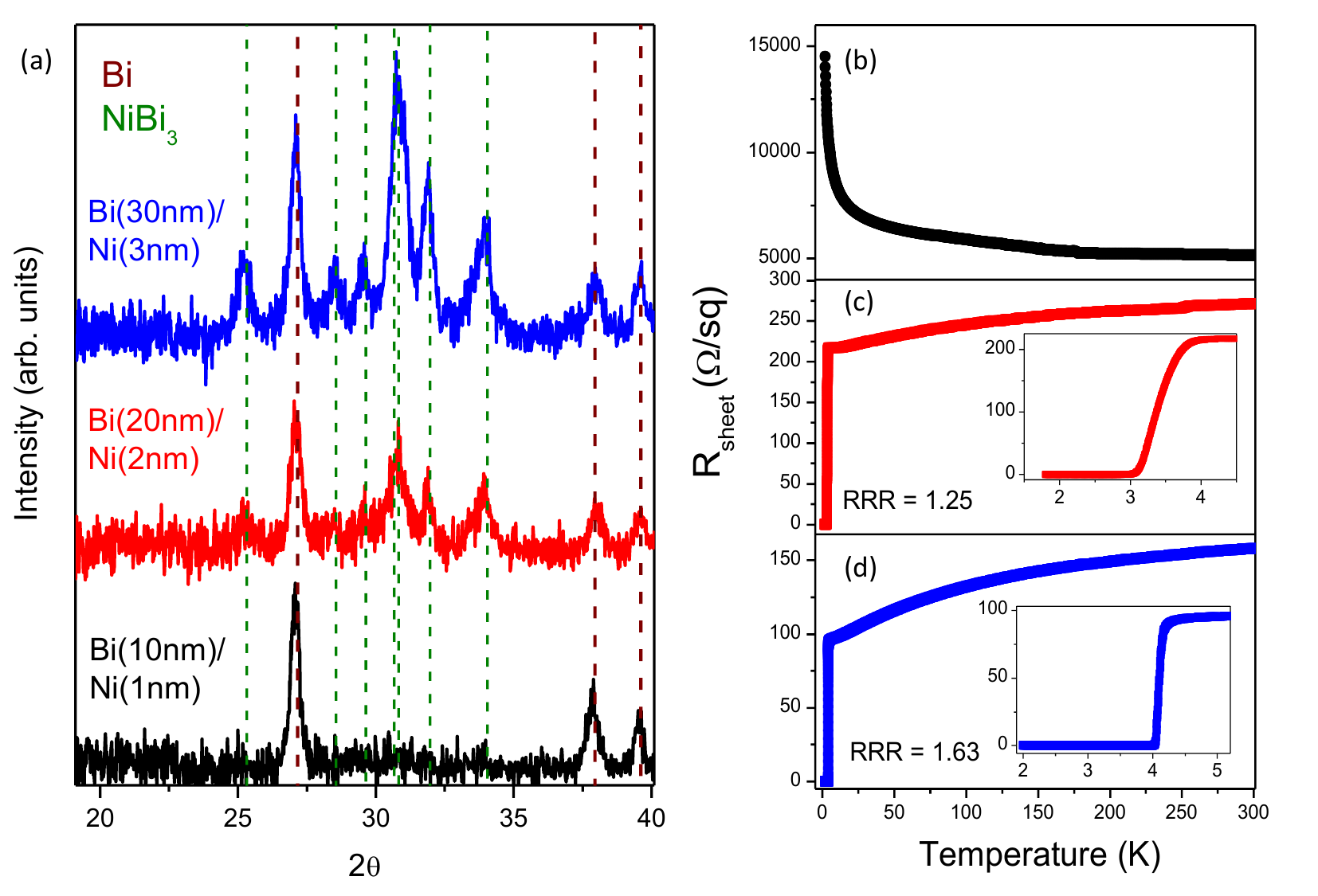}
    \caption{(a) GIXRD diffractogram of Bi(30)nm/Ni(3nm), Bi(20nm)/Ni(2nm) and Bi(30nm)/Ni(3nm). Wine and Green dashed lines stems for Bi rhombohedral and NiBi$_3$ orthorhombic structure index, respectively. Sheet resistance in function of temperature for (b)  Bi(10nm)/Ni(1nm) (c)  Bi(20nm)/Ni(2nm) and (d)  Bi(30nm)/Ni(3nm). \textit{Inset:} zoom-in of the superconducting transition. }
    \label{fig:xrd_rxt}
\end{figure*}

Bi(10nm)/Ni(1nm), Bi(20nm)/Ni(2nm), and Bi(30nm)/Ni(3nm) bilayers were deposited on Al$_2$O$_3$ substrates using an AJA Orion 8 magnetron sputtering system. The base pressure was maintained at 3×10$^{-8}$ Torr, and the working pressure during deposition was 2 $m$Torr in an argon atmosphere. The Ni layers were grown using a DC source at 30 W, with a deposition rate of 0.15 \textup{~\AA}/s, while Bi was deposited using an RF source at 10 W and a rate of 0.20 \textup{~\AA}/s. Fig. \ref{fig:xrd_rxt}{\color{blue}a} presents Grazing-Incidence X-Ray Diffraction (GIXRD) patterns measured at an incident angle of 1$^\circ$. For the 20 nm and 30 nm Bi samples, Bragg peaks associated with both rhombohedral (Bi) and orthorhombic (NiBi$_3$) phases were detected and are indicated by wine and green dashed lines, respectively. In contrast, no signatures of NiBi$_3$ were observed in the 10 nm Bi sample. This absence is consistent with the lack of superconductivity, as shown in Fig. \ref{fig:xrd_rxt}{\color{blue}b}. Fig. \ref{fig:xrd_rxt}{\color{blue}c,d} show the temperature-dependent sheet resistance for the 20 nm and 30 nm Bi samples, both exhibiting metallic behavior resembling of bulk NiBi$_3$, as discussed previously. The insets highlight the superconducting transitions. The Bi(20 nm)/Ni(2nm) sample displays a broad transition with $T_c$ = 3.53 K and $\Delta T_c \approx$ 0.9 K, while the Bi(30nm)/Ni(3nm) sample shows a sharper transition with $T_c$ = 4.08 K and $\Delta T_c$ = 0.1 K.

\begin{figure*}[h]
    \centering
    \includegraphics[width=1.0\linewidth]{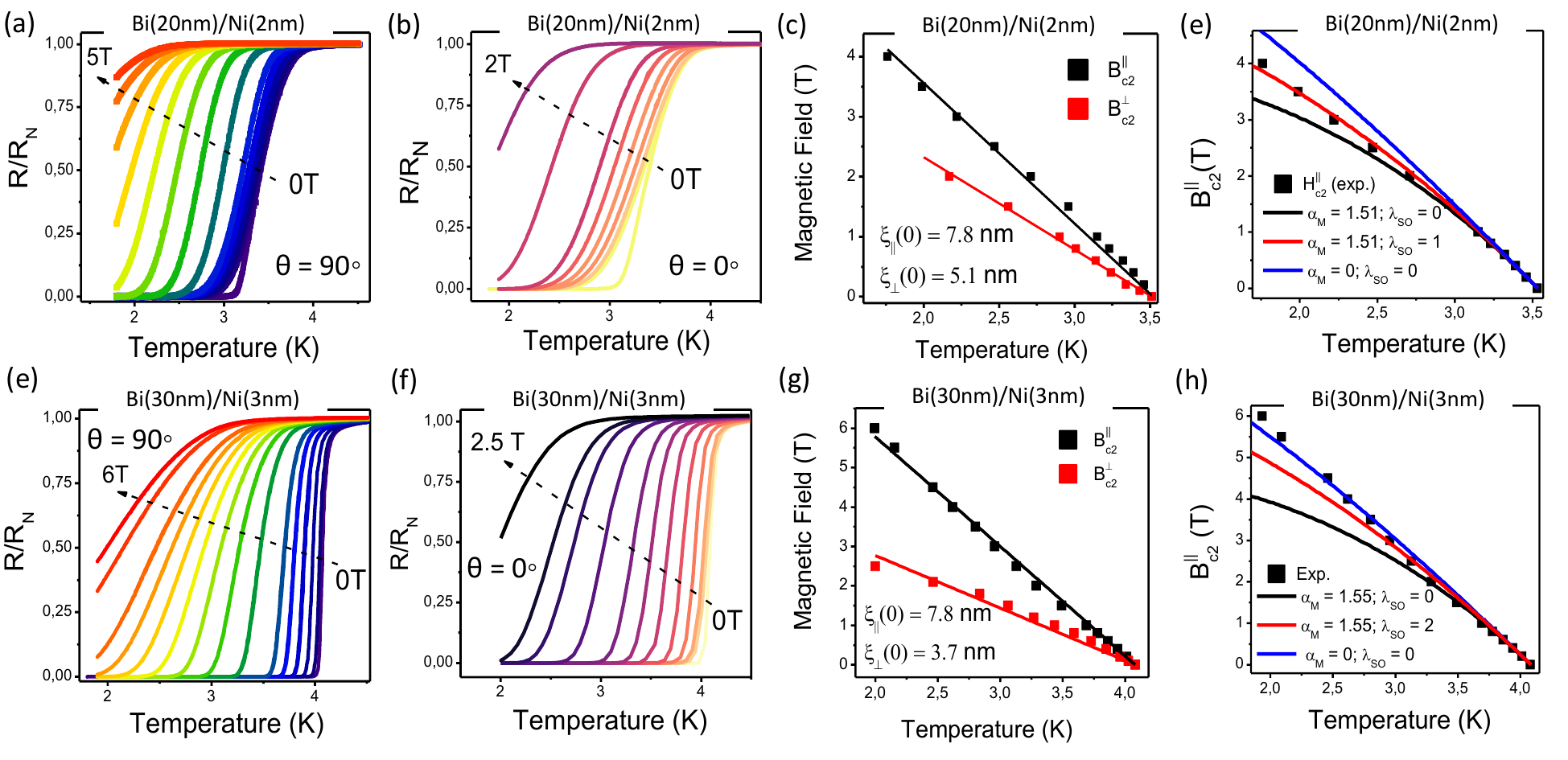}
    \caption{Normalized resistance in function of temperature for Bi(20 nm)/Ni(2 nm) with (a) in plane and (b) out-of-plane magnetic field. The same measurements with respectively magnetic field direction (d) and (c) for Bi(30 nm)/Ni (3 nm). (c),(g) B-T phase diagram. Solid lines stems from the AGL equations.(e),(h) Parallel upper-critical field in function of temperature. Solid colored- lines refers to solution of Eq. \ref{eqn:whh} for a set of parameters - on the panel.}
    \label{fig:sputteringsamples}
\end{figure*}

Magnetotransport measurements with in-plane (Fig. \ref{fig:sputteringsamples}{\color{blue}a,e}) and out-of-plane (Fig. \ref{fig:sputteringsamples}{\color{blue}b,f}) magnetic field for Bi(20 nm)/Ni(2 nm) and Bi(30 nm)/Ni(3 nm) respectively revels a small anisotropy.  Likewise, the B$_{c2}$ defined as 50\% of normal resistance were plot in function of temperature are plotted in Fig. \ref{fig:sputteringsamples}{\color{blue}c,g}. The vortex phase was analyzed with Anisotropy Ginzburg Landau (AGL) model obtaining $\xi_{||} \approx$ 8 nm for both sample, whose value are similar to MBE-grown samples. For the in-plane magnetic field configuration, we applied the 3D AGL model, as the data exhibit a linear trend. The best fit yields a perpendicular coherence length of $\xi_{\perp} \approx 5.1$ nm for Bi(20nm) and $\xi_{\perp} \approx 3.7$ nm for Bi(30nm).

To investigate microscopically the B$_{C2}^{||}$ data, we analyzed it using and Werthamer-Helfand-Hohenberg (WHH) \cite{werthamer1966temperature}. In this model, orbital and paramagnetic effect are included and quantified through the Maki parameter $\alpha$ = $\sqrt{2}B_{c2}^{orb.}/Bp$, where B$_{c2}^{orb.}(0)$ is the orbital limiting field at zero temperature, estimated in the dirty limit as B$_{c2}^{orb.}(0) = -0.693\cdot T_{c}\cdot |dBc2/dT|_{T=T_C}$. Bp $= 1.86\cdot T_c$ is the paramagnetic limit \cite{clogston1962upper}. Spin-orbit scattering (SOS) is incorporate by the dimensionless parameter $\lambda_{SO} = \hbar/(3\pi k_{B}T_{c}\tau_{SO})$, where $\tau_{SO}$ is the SOS lifetime. For a 3D BCS superconductor, the temperature evolution of B$_{c2}$ is given by 

\begin{widetext}
\begin{align}
 \ln\left(\frac{1}{t}\right) = \sum_{v = -\infty}^{\infty} \bigg(\frac{1}{|2v + 1|} -\bigg[|2v + 1| + \frac{\bar{h}}{t} + \frac{(\alpha_{M}\bar{h}/t)^{2}}{|2v + 1| + (\bar{h} + \lambda_{SO})/t}\bigg]\bigg),
  \quad\mbox{where}\quad
  \bar{h} = \frac{4}{\pi^{2}}\frac{B_\mathrm{c2}(t)}{|dB_\mathrm{c2}(t)/dt|}_{t = 1}.
  \label{eqn:whh}
\end{align}    
\end{widetext}
 
From the slope close T$_C$ of Fig. \ref{fig:sputteringsamples}{\color{blue}c,g}, we calculate B$_{C2}^{orb.}$(0) $\approx$ 7.01 T for Bi(20nm) and 8.32 T for Bi(30nm). Likewise, B$_P$ is about . Therefore, $\alpha \approx $ 1.51 and 1.55, respectively. In Fig. \ref{fig:xrd_rxt}{\color{blue}d,h}, $B_{c2}^{\parallel}$ in function of temperature with different WHH fit is shown.  The best fit for Bi(20m)/Ni(2nm) accounts $\alpha $ = 1.51 and $\lambda_{SO}$ = 1, yielding to $\tau_{SO}$ $\approx 230$ fs. As for Bi(30nm)/Ni(3nm), the best fit is the bulk limit case with $\alpha_M = \lambda_{SO}$ = 0. Therefore, for 3D NiBI$_3$ layer, the inclusion of spin-orbit scattering (SOS) 

\section{Beyond B$_P$: Spin-Orbit Scattering}

In the dirty limit for quasi-2D superconductor, the competition between spin paramagnetism and the enhancement due to disorder-induced spin-orbit scattering (SOS) is incorporated in the Klemm-Beasley-Luther (KLB) \cite{klemm1975theory} theory, which describe the temperature dependence of H$_{C2}$ as
\begin{equation}
    ln\left(\frac{T}{T_C}\right) = \psi\left(\frac{1}{2}\right) - \psi\left(\frac{1}{2} + \frac{3\tau_{SO}(\mu_{B}H_{c2})^{2}}{4\hbar\pi k_{B}T}\right)  \label{eqn:klb},
\end{equation}
where $\psi(x)$ is the digamma function and $\tau_{SO}$ represents the spin-orbit scattering lifetime.

\begin{figure*}
    \centering
    \includegraphics[width=0.5\linewidth]{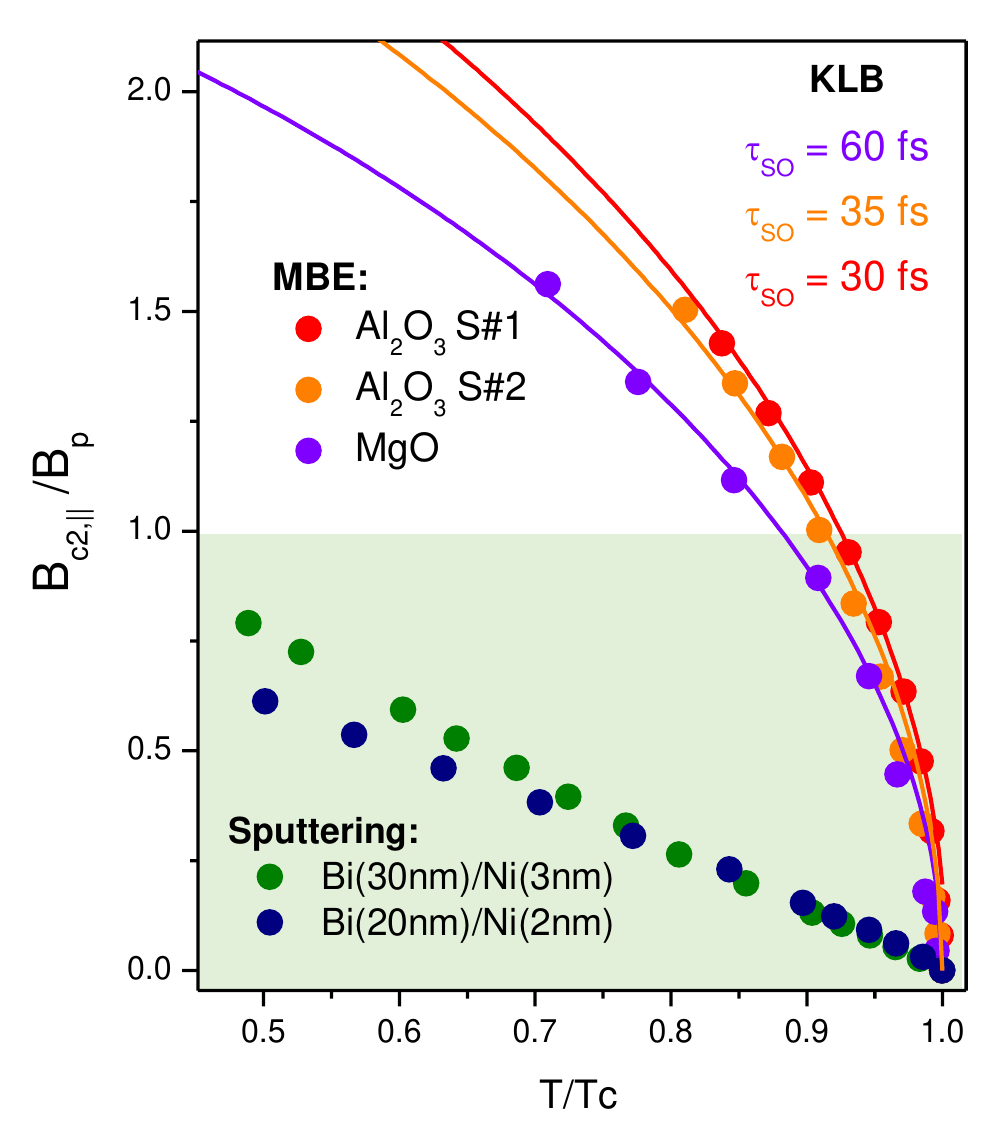}
    \caption{Normalized In-plane upper critical field divided by B$_P$ (green shaded region) as function of reduced temperature for S$\#$1,S$\#$2, MgO and the sputtering samples Bi(10$\cdot$ x nm)/Ni(x nm) with x=2 and 3. Colored lines represents the best fit of KLB model (Eq. \ref{eqn:klb}) along with the extracted spin-orbit scattering lifetime ($t_{SO}$).}
    \label{fig:KLB_reducedPD}
\end{figure*}

Similar to the main text, Fig.~\ref{fig:KLB_reducedPD} presents the reduced phase diagram of $B_{c2}/B_{P}$ as a function of reduced temperature. The data shown correspond to both the 
MBE-grown samples and the sputtered film, the latter included for comparison since it follows a BCS-like behavior and does not violate the spin-paramagnetic limit. As indicated by the 
colored lines, the best fits to Eq. \ref{eqn:klb} yield spin-orbit scattering times of $\tau_{\mathrm{SO}} = 30$ fs for S\#1, $\tau_{\mathrm{SO}} = 35$ fs for S\#2, and $\tau_{\mathrm{SO}} = 60$~fs for the MgO sample.

Nevertheless, the KLB model relies on Matthiessen's rule, and therefore $\tau_{\mathrm{SO}}$ cannot be smaller than the total scattering lifetime ($\tau_{\mathrm{tot}}$) \cite{klemm1975theory,liu2018interface}. To verify this condition, we estimate $\tau_{\mathrm{tot}}$ using the Drude expression $\tau_{\mathrm{tot}} = m^{*}/(n e^{2} \rho)$, where $\rho$ is the longitudinal resistivity and $m^{*}$ is the effective mass. The carrier density of NiBi$_3$ at 5~K was found to be $n = 3.2\times 10^{27}\,\mathrm{m^{-3}}$ from transverse measurements (see main text) for S\#2, consistent with Ref. \cite{sant2026linear}. Using $m^{*} = 12 m_{e}$, as reported in Ref. \cite{fujimori2000superconducting} for polycrystalline NiBi$_3$, and $\rho \approx R_{\mathrm{sheet}}\, t \approx 95~\mu\Omega\,$cm, we obtain $\tau_{\mathrm{tot}} \approx 140$~fs at 5 K for S\#2. This value exceeds the extracted $\tau_{\mathrm{SO}}$ by a factor of 4, contradicting a fundamental requirement of the theory ($\tau_{\mathrm{SO}} \gg \tau_{\mathrm{tot}}$) \cite{lu2015evidence,saito2016superconductivity}. Based on this, we conclude that disorder-induced spin-flip scattering is unlikely to be responsible for the enhancement of $B_{c2}^{\parallel}$ beyond the paramagnetic limit in 2D NiBi$_3$.

\bibliography{apssamp}

\end{document}